\documentclass[aps,prl,twocolumn,groupedaddress,showpacs]{revtex4-1}

\usepackage{color}
\usepackage{amsmath}
\usepackage{graphicx}
\usepackage{bm}
\usepackage{bbm}
\usepackage[normalem]{ulem}
\usepackage[geometry]{ifsym}
\usepackage{latexsym}
\usepackage{wasysym}

\begin{document}


\title{Feeding ducks, bacterial chemotaxis, and the Gini index}
\author{Fran\c{c}ois J. Peaudecerf and Raymond E. Goldstein}
\affiliation{Department of Applied Mathematics and Theoretical Physics, Centre for 
Mathematical Sciences,\\ University of Cambridge, Wilberforce Road, Cambridge CB3 0WA, 
United Kingdom}

\date{\today}

\begin{abstract}
Classic experiments on the distribution of ducks around separated food sources found consistency with
the `ideal free' distribution in which the local population is proportional to the local supply rate.  
Motivated by this experiment and others, we examine the analogous problem in the microbial world: the 
distribution of chemotactic 
bacteria around multiple nearby food sources. In contrast to the optimization of uptake rate that may hold at
the level of a single cell in a spatially varying nutrient field, nutrient consumption by a population of 
chemotactic cells will modify the nutrient field, and the uptake rate will generally vary throughout the population. 
Through a simple model we study the distribution of resource uptake in the presence of chemotaxis, consumption, 
and diffusion of both bacteria and nutrients. Borrowing from the field of theoretical economics, we
explore how the Gini index can be used as a means to quantify the inequalities of uptake.  The 
redistributive effect of chemotaxis can lead to a phenomenon we term `chemotactic levelling',
and the influence of these results on population fitness are briefly considered.
\end{abstract}


\maketitle

\section{Introduction}

In one of the more amusing, yet influential experiments on animal behavior, Harper \cite{ducks} 
studied the distribution of mallards around two separated sources of standardized pieces of bread. 
After an induction period on the order of a minute, the average number of ducks clustered tightly 
around each station stabilized.
The distribution he observed was simple: the number of ducks at each source was 
proportional to the flux of bread there (pieces/minute). This constituted the first
experimental observation of the so-called \emph{ideal free distribution} previously introduced
in theoretical ecology.  Using the terminology of 
Fretwell and Lucas \cite{FretwellLucas}, `ideal' means that ducks can identify the source where 
their uptake is maximized, and `free' implies unfettered ability to access the source of choice.  
This distribution, resulting from individual rational behaviors, achieves a population-wide 
uniformization of the probability of uptake, and can be understood as a Nash equilibrium \cite{Nash}. 
These works impacted not only 
ornithology, but ecology \cite{ecology1, ecology2}, evolutionary biology \cite{evolution} and the study of 
human behavior \cite{human1, human2, human3}, all areas involving
resource acquisition in a heterogeneous environment.

Here we take motivation from Harper's experiment, and others discussed below, to examine resource 
acquisition in a heterogeneous \emph{microbial} world \cite{Azam} where swimming microorganisms 
respond to nutrient sources through concentration fields determined by molecular diffusion and microbial uptake.  
For the specific case of peritrichously flagellated bacteria such as {\it E. coli} and {\it B. subtilis}, 
cells move in a run-and-tumble random walk biased by  
concentration gradients, resulting in drift of the population up these gradients \cite{Berg}. 
Because chemotaxis 
\cite{BergPurcell} is quite different from the visually-based searching of 
higher animals, and because of the diffusive behavior of nutrients and the cell populations  
the microbial problem is distinct in character. This feature motivates the present investigation of the 
consequences of a collection of individual chemotactic responses on the population-scale distribution of resources.  
While chemotaxis is generally thought to optimize uptake at the single-cell level \cite{Celani}, 
even the mere presence of translational diffusion in a population would lead to a distribution of uptake rates.
And the interplay of chemotaxis and consumption will modify the distribution of resources and cells, with
further potential impact on the uptake rate distribution.  In this paper we focus on three key questions in this
area: 
What is the distribution of bacteria around spatially distinct nutrient sources and their associated impact 
on the resource field?  
What is the distribution of resource uptake rates within that population? 
What are the consequences of such distributions for cellular fitness?

A historically important experiment on spatially-varying resources is Engelmann's 1883 determination of
the action spectrum of photosynthesis \cite{Engelmann_photosynthesis}.  
Having discovered bacteria that are attracted to the oxygen produced by 
photosynthesis \cite{Engelmann_oxygentaxis},
he imaged the solar spectrum onto a linear algal cell in an air-tight chamber containing such bacteria.  
They clustered around the alga in proportion to the local oxygen production, 
revealing with greater precision than the available techniques of the time the peaks of 
photosynthetic activity for blue and red wavelengths. How reliably the local bacterial accumulation 
reflects the oxygen production rate, in the face of both bacterial and oxygen diffusion, remains an 
open question in the spirit of the present investigation. 
Moreover, this work demonstrates how micro-domains releasing a limited quantity of attractive 
nutrients in a continuous fashion can arise in the microbial world.

Understanding bacterial organization and uptake around algal resources also finds 
an important biological context in the case of 
\emph{bacterial-algal symbiosis}. The recent discovery \cite{Croft} that many algae dependent 
on vitamin B12 obtain it through symbiotic relationships with bacteria raises 
questions about spatio-temporal aspects of symbiosis: how the two species find each other and arrange 
themselves to achieve the symbiosis. 

These are examples of a more general problem of microorganisms 
responding to the `patchy' nature
of nutrients in ecosystems \cite{Blackburn}, such as the prosaically named
`marine snow' \cite{snow}. We emphasize the fundamental difference between live microbial sources, 
such as Engelmann's alga, and inert sources, such as lysis events; the former 
continuously release nutrients at low rates. They are thus stable in time but can be significantly 
impacted by bacterial uptake. These characteristics makes them the natural microbial equivalent to 
the limited continuous sources considered by Harper in his animal experiments.
 
To make concrete the interplay between production, consumption, 
diffusion, and chemotaxis we consider a generalized Keller--Segel (KS) model \cite{KellerSegel}.
While the KS model is a well-established model that has found frequent application in the study of spatially
extended microorganism populations \cite{Murraybook}, the biological setting of localized, low-intensity 
sources with a steady release of nutrients is little-studied, and the overarching issue of resource
uptake distribution is essentially unexplored.  In what follows the steady-state 
distributions arising from multiple localized sources are analyzed to understand the consequences of chemotaxis 
on the distribution of uptake in the bacterial population, the total uptake being fixed.  Borrowing from 
theoretical economics, we next propose that the Gini index \cite{Gini}, a number originally used to characterize 
wealth inequality, can be used to quantify the individual uptake distribution. 
Varying the model parameters and  the dimensionality of space we show that chemotaxis can switch from 
redistributing the resource to generating greater inequalities. Finally, 
we explore the potential biological consequences of uptake redistribution through 
an example of growth at low nutrient levels.

\section{The Model}

\begin{figure}[t]
\includegraphics{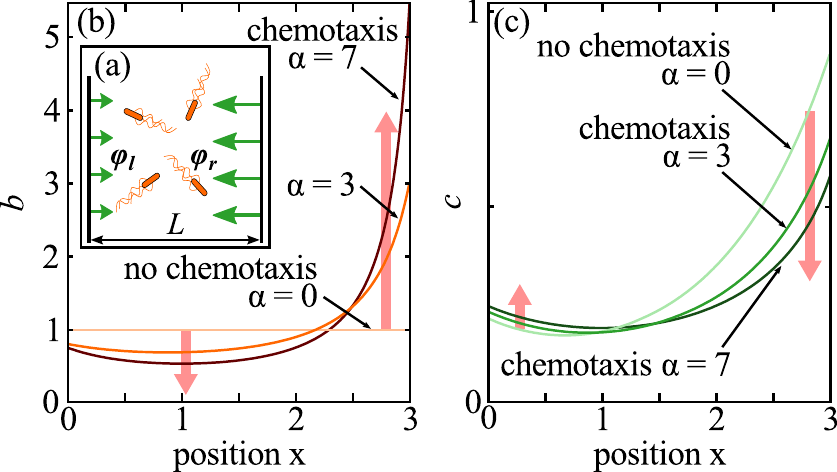}
\caption{(color online). Numerical results in one dimension. (a) Sketch of the setup. (b) Example of a steady-state distribution 
of bacteria in the rescaled KS model \eqref{consumption_nodim} \& \eqref{chemotaxis_nodim}. (c) Corresponding distribution of nutrient 
at steady state. In (b) \& (c) the parameters values are \(\lambda=3\)  for the domain size parameter and \(s=1/8\) for the relative strength of the left source. The importance of chemotaxis against bacterial diffusion increases with the chemotactic parameter \(\alpha\). }
\label{fig1}
\end{figure}

Consider bacteria with mean concentration $b_0$ and local concentration $b({\bf r},t)$ 
in a $d$-dimensional volume $L^d$.  Within the volume are 
nutrient sources with fluxes $\{\phi_i\}$ of typical value $\phi_0$ leading to a concentration field $c({\bf r},t)$. 
$b$ and $c$ obey KS equations \cite{KellerSegel},
\begin{eqnarray}
\frac{\partial c}{\partial t}&=&D_c\nabla^2 c-bf(c)~,\\ \label{consumption}
\frac{\partial b}{\partial t}&=&D_b\nabla^2 b -\bm{\nabla}\cdot\left[\chi(c) b \bm{\nabla}c\right]~,
\label{chemotaxis}
\end{eqnarray}
with $D_c$ and $D_b$ the diffusion coefficients of nutrients and bacteria respectively, and 
$\chi(c)$ the chemotactic response coefficient. 
The nutrient uptake rate $f(c)$ per bacterium is expected to behave at low $c$ as $f(c)\sim kc$, 
with saturation at high $c$: $f(c)\sim k_{max}$. When the nutrient 
is essential for life, such as oxygen for obligate aerobes, $f(c)$ may  
vanish below some 
$c^*$ \cite{Hillesdon,pnas_swim}.  As in Harper's study and Engelmann's experiment, the interesting 
regime has the resource limiting, with $b_0$ and 
$L$ such that the total nutrient flux can be consumed and is thus below the maximum value
$k_{max} b_0 L^d$.  Otherwise, 
steady state cannot be attained and $c$ increases indefinitely.

For small $c$, we identify the length scale $\ell_k=(D_c/kb_0)^{1/2}$ for
concentration gradients due to uptake.  For the a run-and-tumble chemotaxis mechanism to operate 
and the continuum model to be relevant, $\ell_k$
should be large compared to the run length $\ell_{run}=v\tau$, where $v$ is a typical swimming speed and $\tau$ the time
between tumbles. That is, we require the Knudsen-like number $l_{run}/l_k\ll 1$.  For {\it E. coli}, with 
$v\sim 20$ $\mu$m/s and $\tau\sim 1$ s, $\ell_{run}\sim 20$ $\mu$m, but it can
be considerably longer for other bacteria \cite{other_bacteria}.  Still at low $c$, with 
$\chi(c)=\chi_0$ \cite{Tindall}, 
interesting behavior occurs when the 
chemotactic flux $\sim \chi_0 b_0 \phi_0 / D_c$ 
dominates the diffusive flux $\sim D_b b_0 /\ell_k$,
with $\phi_0/D_c$ a typical concentration gradient.  Thus, the
P\'eclet-like number 
\begin{equation}
\alpha=\frac{\chi_0\phi_0}{D_bD_c\ell_k}
\label{Peclet}
\end{equation}
exceeds unity.  

A one-dimensional version of Harper's experiment (Fig. \ref{fig1}(a)) has fluxes 
$\phi_l$ and $\phi_r$ at the left and right domain boundaries, and $\Phi=\phi_l+\phi_r$.  
Nondimensionalizing as
above we obtain in the low-$c$ regime
\begin{eqnarray}
\frac{\partial c}{\partial t}&=&\nabla^2 c-bc~,\label{consumption_nodim} \\ 
\delta^{-1}\frac{\partial b}{\partial t}&=&\nabla^2 b -\alpha\bm{\nabla}\cdot\left[ b \bm{\nabla}c\right]~,
\label{chemotaxis_nodim}
\end{eqnarray}
where $\delta=D_b/D_c$, with the following nondimensional boundary conditions:
\begin{eqnarray}
\left.{\bf \hat{n}} \cdot {\bm \nabla} c \right|_{0} =& - s  \;\;\;\;\;\;\;\;{\bf \hat{n}} \cdot \left.\left({\bm \nabla} b - \alpha b {\bm \nabla} c\right)\right|_{0} =& 0\\
\left.{\bf \hat{n}} \cdot {\bm \nabla} c \right|_{\lambda} =& \;1 - s \;\;\;\;\;\;\;{\bf \hat{n}} \cdot \left.\left({\bm \nabla} b - \alpha b {\bm \nabla} c\right)\right|_{\lambda} =& 0~,
\end{eqnarray}
where ${\bf \hat{n}}$ is the outward unit normal to the domain.

Two new parameters drive the evolution of the system: $\lambda=L/\ell_k$, the domain 
size relative 
to the screening length, and the relative strength $s=\phi_l/\Phi$ of the left source. 
Before investigating the steady-state solutions to \eqref{consumption_nodim} and \eqref{chemotaxis_nodim}, 
we point out that by construction 
the total uptake \(U\) over the population is equal at steady-state to the total nutrient flux into 
the chamber,
\begin{equation}
U = \int_0^{\lambda}\! dx\, b c\; =\int_0^{\lambda}\! dx \nabla^2 C 
= \left.{\bf \hat{n}}\cdot {\bm \nabla} c\right|_0^{\lambda}  = 1~, 
\end{equation}
independent of the choice of parameters and strength of chemotaxis. 
Thus overall uptake optimization is not part of the present study: our interest resides instead in how the 
individual behaviors result in a distribution of finite resource among population members, 
just as in Harper's experiment.

One of the natural questions to ask is whether there is a variational structure to the KS equations \eqref{consumption_nodim}
and \eqref{chemotaxis_nodim}.   The only related result of which we are aware concerns the case when the nutrient 
consumption rate $f(c)$ takes the aforementioned high-$c$ form of a constant.  After suitable rescaling that dynamics is 
\begin{eqnarray}
\frac{\partial c}{\partial t}&=&\nabla^2 c+\nu b~,\label{consumption_nodim1} \\ 
\delta^{-1}\frac{\partial b}{\partial t}&=&\nabla^2 b -\alpha\bm{\nabla}\cdot\left[ b \bm{\nabla}c\right]~,
\label{chemotaxis_nodim1}
\end{eqnarray}
where $\nu=-1$.  
As shown recently \cite{KS_variational}, a dynamics of a closely related form is variational. If we introduce 
the energy functional
\begin{equation}
{\cal E}[b,c]=\int\! d{\bf x} \left\{ \frac{1}{\alpha}b\log b -bc +\frac{1}{2}\vert{\bm \nabla} c \vert^2 \right\}
\label{Efunc}
\end{equation}
then the variational relations
\begin{equation}
\frac{\partial c}{\partial t}=-\frac{\delta {\cal E}}{\delta c}~,\ \ \ \ \delta^{-1}\frac{\partial b}{\partial t}=
\alpha {\bm \nabla}\cdot \left[ b{\bm \nabla}\frac{\delta {\cal E}}{\delta b}\right]~.
\end{equation}
yield \eqref{consumption_nodim1} and \eqref{chemotaxis_nodim1}, but with $\nu=+1$.  This case corresponds to the 
well-studied situation in which the bacteria are sources for the chemoattractant, rather than sinks.  Because the same term 
($-bc$) in ${\cal E}$ yields both the production/consumption term in \eqref{consumption_nodim1} and the chemotactic term
in \eqref{chemotaxis_nodim1}, the case $\nu=-1$ appears
not to have any variational structure of this type \cite{Wolansky}.   But since the consumption 
rate per bacterium is constant in
this regime, the 
distribution of resource acquisition rates is trivial and
a variational structure would provide no new information.  More importantly, the case under consideration here, with $f(c)\sim c$ also appears not to possess a
variational structure and thus it is not possible to conclude that the steady-state solutions are in 
any way minimizers or maximizers of some energy-like functional.  It follows that the distribution of uptake rates is a nontrivial
feature of the underlying diffusion-consumption-chemotaxis dynamics.

Figures \ref{fig1}(b) and \ref{fig1}(c) show steady state distributions of $b$ and $c$ for $\alpha=3$, 
$\lambda=3$, $\delta=0.6$ and $s=1/8$, which correspond to physical values of the dimensional parameters 
($D_c=5\times 10^{-6} \:\text{cm}^2 \,\text{s}^{-1}$, $D_b=3\times 10^{-6} \:\text{cm}^2 \,\text{s}^{-1}$, 
$k=2\times 10^{-7} \:\text{cell}^{-1} \,\text{s}^{-1} \,\text{cm}^3$, $\chi_0=2.7\times 10^{-3}  \:\text{cm}^2 \,\text{s}^{-1} 
\,\text{mM}^{-1}$, $b_0=10^6 \:\text{cells} \,\text{cm}^{-3}$, $\phi_0=3.4\times 10^{-6} \:\text{mM} \,\text{cm} \,\text{s}^{-1}$ and 
$L=150 \:\mu$m \cite{Natarajan, Ahmed}), along with the non-chemotactic case and the case $\alpha=7$. 
Bacteria accumulate on both sides, with more 
closer to the stronger source, as one might expect. Intriguingly, this leads to what we term
{\it chemotactic levelling} of the nutrient: a more uniform 
concentration 
field than without chemotaxis.  In particular, we notice that the maximum uptake rate of a 
bacterium in this population, obtained closest to the strongest source, is decreased by chemotaxis. 
Diffusion of $b$ and $c$ precludes
the ideal free distribution. 

\section{Quantifying Inequalities with the Gini Index}

Whereas studies on biological consequences of chemotaxis usually measure the increased uptake over the whole or 
part of the population with respect to the non-chemotactic case (e.g. \cite{Stocker,Taylor}), as emphasized above,
the present study is focused on the case for which the mean uptake rate is independent of the chemotactic behavior. 
As our interest then resides in how equally resources are spread among the population,
comparison of the chemotactic results with the non-chemotactic reference distributions 
thus requires a measure of the proximity with the ideal free distribution. 
Among the many possible measures of inequality \cite{Cowell}, we consider here the Gini index $G\in [0,1]$ \cite{Gini}, 
which for a distribution $P(w)$ of wealth $w$ in 
a population can be expressed as
\begin{equation}
G=\frac{\int\!\!\int\! du dv P(u) P(v)|u-v|}{2\int\! du\, u P(u)}.
\label{Gini}
\end{equation}
The ideal free distribution, in which every individual has the same wealth $u_0$, is
$P=\delta(u-u_0)$ and thus $G=0$, while larger values hold for more unequal distributions. $G$ can be 
used with any notion of `wealth' \cite{Gini2}, such as biodiversity \cite{Wittebolle}.
We should empasize that in using the Gini index to quantify uptake inequalities in the present study we do 
not imply any preferred status to $G$ as a metric for resource acquisition distributions.  There are many that
could be explored, the Gini index presenting the advantage of being easily translated in the context of the continuous distribution and thus enabling analytical work on its variations.
Using the individual nutrient uptake rate as wealth, we can transform the integrals in the uptake rate levels into integrals in space. This makes the bacterial density appear as the equivalent in space of the frequency distribution of uptake levels. Normalizing this new density function, thus making the integral for the total number of bacterial cells appear at the denominator, we finally re-express \eqref{Gini} for 
time-dependent spatial distributions in a domain $\Omega$:
\begin{equation}
G(t)=\frac{\int_{\Omega} \int_{\Omega}dx dy\,b({\bf x},t) b({\bf y},t)|f(c({\bf x},t))-f(c({\bf y},t))|}
{2 \int_{\Omega}dx\, b({\bf x},t) f(c({\bf x},t)) \int_{\Omega}dx\, b({\bf x},t)}.
\label{Gini_bact}
\end{equation}

\begin{figure}[t]
\includegraphics{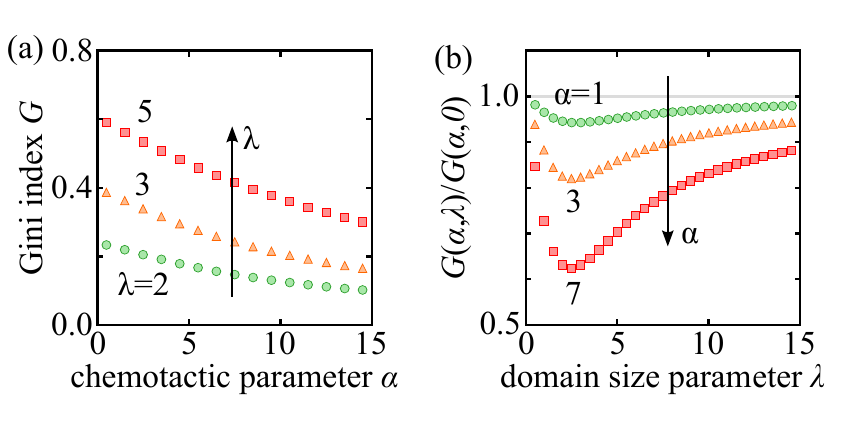}
\caption{(color online). Variations of Gini index for bacterial uptake, with $s=0$ (single source on the right side). (a) $G$ at steady-state with respect to the chemotactic 
parameter \(\alpha\) for various domain sizes $\lambda$. (b) Ratio of $G$ in the chemotactic case to that without, 
for various chemotactic strengths $\alpha$, showing optimal domain size parameter for relative decrease of $G$.}
\label{fig2}
\end{figure}

With this measure of inequality, we investigate how the uptake distribution at steady-state depends on the parameters of
the KS model \eqref{consumption_nodim}-\eqref{chemotaxis_nodim}. 
 Before proceeding we should make clear that the Gini index values calculated within the present approach and the corresponding inequalities in the uptake levels 
 are instantaneous:  physically, bacteria would swim along a biased random walk inside the steady-state nutrient 
 distribution, thus sampling different concentration levels. Over a time long in comparison with the typical time of 
 bacterial diffusion at the scale of the experimental chamber this motility would tend to level the integrated uptake 
 within the population and yield lower values of $G$.  The fundamental issue is then whether the time scale
 for this smoothing-out of inequalities is large or small compared to the timescale \(\tau_{int}\) for a relevant 
 internal biological process based on nutrient uptake. The present approach is thus in the limit
 \(\tau_{int} D_b / L^2 \ll 1\), thus most relevant to large system sizes and short internal times.
As an example, it takes approximately 17h for a typical run-and-tumble bacteria (\(D_b = 4. 10^{-6} \;\mbox{cm}^2.\mbox{s}^{-1}\)) to explore the space between two sources separated by 5 mm, a time that is much bigger than the typical scale of key cellular processes such as division (approx. 30 minutes).
 
We now go back to investigating the role of each parameter: the ratio of diffusion coefficients $\delta$ impacts 
transient dynamics but does not modify steady state solutions. From numerical solutions in the 
phase space delimited by $s \in [0,0.5]$ (from one source on the right to equal sources), $\alpha \in [0,15]$ (strength of chemotaxis) 
and $\lambda \in[0.5,15]$ (domain size), we obtain first the intuitive result that for given chemotactic and domain size 
parameters, the more equal the sources are, the more equal the uptake is among the population, and the lower is $G$. More 
balanced sources indeed create smaller nutrient gradients, thus a lesser range of uptakes and weaker chemotaxis.
We also find that $G$ increases with the size parameter $\lambda$ (Fig. \ref{fig2} (a) in the case $s=0$),
for larger $\lambda$ corresponds to stronger variations of the concentration field and higher variations of uptake. 

\begin{figure}[t]
\includegraphics{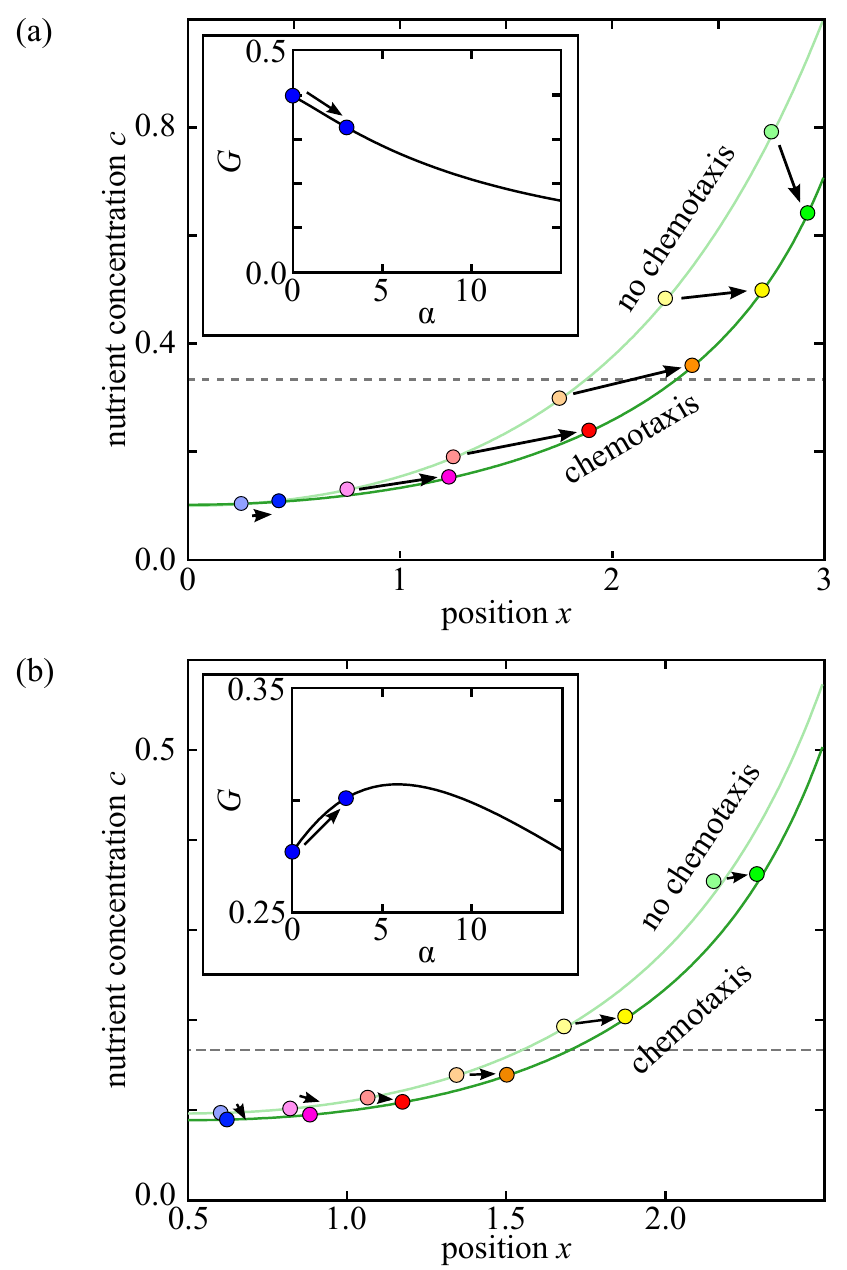}
\caption{(color online). Interpretation of steady-state Gini index variation in different dimensions $d$. 
(a) $d=1$ with domain size parameter $\lambda=3$, $s=0$ (single source on right side), chemotactic strength parameter $\alpha=3$.  (b) $d=2$ with a circular source of size $l=0.5$ in a domain of radius $L=4.5$, and chemotactic strength parameter $\alpha=3$.
Shown are center of mass and 
mean uptake level of the sextiles of uptake distribution (colored circles), overlaid on nutrient concentration, 
without and with chemotaxis. Dashed line shows concentration  
for average uptake. Inset: $G$ vs. chemotactic strength parameter $\alpha$, with blue circles corresponding to  
displayed solutions. }
\label{fig3}
\end{figure}

The question of whether chemotactic levelling of the nutrient concentration field can make the distribution of individual uptake  
more ideal is addressed by varying $\alpha$. Its impact 
is subtle: chemotaxis levels nutrients across the domain, 
but accumulation of cells near sources improves uptake of some to the detriment of others. 
We find that $G$ actually decreases with $\alpha$ (Fig. \ref{fig2}(a) for a single source), 
which reveals chemotactic levelling of the uptake rate among the population. In Fig. \ref{fig1}, $G$
decreases from $\simeq 0.3$ with no chemotaxis to $\simeq 0.25$ when chemotaxis ($\alpha=3$) is allowed.

This decrease is best understood for a single source on 
one side of the domain. Because $c$ decreases monotonically from the source, the 
bacterial population can be split into quantiles of uptake that 
are ordered in space. 
Chemotaxis lowers the nutrient concentration over the whole domain, but mostly close to the source, and it shifts the center 
of mass and mean uptake level of each quantile.  Fig. \ref{fig3}a show that bacteria with the higher uptake, which are 
also closer to the source, are transferred to lower uptake levels. For the lower uptake quantiles, higher levels of uptake 
compared to the non-chemotactic case are attained due to chemotaxis toward the source.  Together, these effects bring uptake 
levels closer to the average, lowering $G$: the bacterial system moves closer to the ideal free distribution.

The generality of this result can be established in the limit of weak
chemotaxis ($\alpha \ll 1$) still with a single source, where a series solution $G \simeq G_0(\lambda)+\alpha \:G_1(\lambda)+\cdots$ of \eqref{consumption_nodim}
and \eqref{chemotaxis_nodim} yields 
\begin{eqnarray}
G_0 (\lambda) &=& 1 - 2\frac{\cosh\lambda-1}{\lambda \sinh\lambda}~,\label{Gini_0}\\
G_1 (\lambda) &=& -\frac{2}{3 \lambda} +\frac{\cosh\lambda-1}{\lambda^2 \sinh\lambda} 
\left(1 + \frac{\lambda \cosh\lambda}{3 \sinh(\lambda)}\right)~,
\label{Gini_1}
\end{eqnarray}
where $G_1(\lambda)<0$, so $G$ indeed decreases with chemotaxis. 
A cumbersome analysis (not shown) for the general case $s \in [0,0.5]$, that is $G \simeq G_0(\lambda,s)+\alpha \:G_1(\lambda,s)+\cdots$, 
also shows that $G_1(\lambda,s)<0$, thus extending this result to any ratio of source stengths. 

In the limit of strong chemotaxis ($\alpha \gg 1$), where bacterial diffusion becomes irrelevant, we may expect to recover the
ideal free distribution.   Analytical progress on this steady-state problem is achieved by integrating twice 
\eqref{chemotaxis_nodim} to obtain $b(c)$, substituting into \eqref{consumption_nodim} and then expanding in powers
of $1/\alpha$.  One obtains
\begin{eqnarray}
b(x) &=& \frac{1}{\alpha}\frac{ \lambda \omega^2}{2 \: \cos^2\left[\omega\left( x-\beta_1\right)/2\right]}\\
c(x) &=& \frac{1}{\lambda}+\frac{1}{\alpha} \left\{\beta_2  -2  \:\mbox{log} \left[ \mbox{cos} \left(\omega (x -\beta_1)/2 \right)\right]\right\}~,
\label{Asympt_1}
\end{eqnarray}
where $\omega$ and $\beta_1$ are determined through
\begin{eqnarray}
\omega\: \mbox{tan}\left(\frac{\omega}{2}\beta_1\right)&=&\alpha s\\
\omega\: \mbox{tan}\left[\frac{\omega}{2}\left(-\beta_1+\lambda\right)\right]&=&\alpha (1-s)~.
\label{Asympt_2}
\end{eqnarray}
The constant $\beta_2$ depends on $\omega$ and the model parameters.  Neglecting terms in $1/\alpha^2$ 
and higher order, it
can be written as
\begin{equation}
\begin{aligned}
\beta_2\simeq & -2\: \mbox{log} \: \alpha + 2 + 2\: \mbox{log} \:\omega \\ & - 2 \left[s\: \mbox{log} \:s + (1-s) 
\:\mbox{log} (1-s) \right] - \omega^2 \lambda / \alpha+\cdots~.
\end{aligned}
\end{equation}

This solution enables us to obtain an analytical expression for the Gini index in the limit of strong chemotaxis: 
\begin{equation}
G \simeq \lambda \left\{1 - 2s + 4s^2+s(1-s)\mbox{log}\left[\frac{(1-s)^2}{s^2}\right]\right\} \frac{1}{\alpha}+\cdots~,
\end{equation}
to leading order in $1/\alpha$, for $s \in [0,0.5]$. This establishes further the generality of its 
decrease with $\alpha$ together with its increase with $\lambda$: in 1D, chemotaxis levels the uptake 
throughout the population. Moreover, in this range of high $\alpha$, inequalities of 
uptake initially increase when the system changes from a single source to more balanced sources.

As this levelling of the uptake distribution appears as the microbial equivalent of the more uniform 
uptake displayed by ducks, it is natural to ask if, as in Harper's experiments, the bacterial population 
reaches the ideal free distribution as $\alpha\to \infty$ and splits 
into two localized sub-populations proportional to the source strengths. If we consider that the position 
\(x_0\) of minimum bacterial concentration separates a left population \(B_L\) associated to the left source 
and its right-hand-side equivalent \(B_R\), our analytical solution for $\alpha \gg 1$ directly 
yields
\begin{equation}
\int_0^{\beta_1}\! dx\, b(x) = B_L \simeq \lambda \;s~.
\end{equation}
Thus, the population associated to one source is directly proportional, to leading order, 
to the flux of this source, as in the ideal free distribution. Moreover, analysis of \eqref{Asympt_1} shows
that in the limit \(\alpha \rightarrow \infty\), \(b(x)\) is localized in regions of width
$\sim 1/\sqrt{\alpha}$ at both \(x=0\) and \(x=\lambda\), with peak values $\sim \alpha$ at these positions.
In the limit of \(\alpha \rightarrow \infty\), we thus get a localisation 
of the number of cells proportional to the source at the source: this is the (unphysical) limit of a microbial 
ideal free distribution.

When the domain size parameter $\lambda\gg 1$, the central 
portion of the domain has a steady-state concentration $c \sim 0$, with very small gradients. Bacteria there are {\it screened} from the sources, unable to feel sufficient gradients to move chemotactically closer 
to them.  The relative redistributive effect of chemotaxis compared to its absence must then diminish with distance. 
Considering the relative Gini index $G(\alpha)/G_0$ in the approximation (\ref{Gini_0})-(\ref{Gini_1}), we indeed 
find an optimal domain size, $\lambda_G \simeq 3.12$, for which the redistributive effect of chemotaxis is the 
strongest. An optimal size 
is also found in simulations beyond the linear regime in $\alpha$ (Fig. \ref{fig2} (b)) 
and with influx from both sides, with $\lambda_G \simeq 3$. The decrease of this relative change 
for high values of  $\lambda$ embodies the aforementioned screening, while the behavior at low $\lambda$ results from
a nearly uniform concentration over the domain, with only weak gradients for a chemotactic response.

Does the uptake levelling found in $d=1$ hold in higher dimensions? To answer this,
we solve \eqref{consumption_nodim} and \eqref{chemotaxis_nodim} for a single spherical source of radius $l$ in a closed spherical domain of radius $L$,
 both measured in units of $l_k$. We find that in 
$d=2$ and $3$ the effects of chemotaxis, for a given size of source and domain, are much weaker. Moreover, for certain 
parameter values, chemotaxis can actually increase $G$ (inset in Fig. \ref{fig3} (b)). Analysis of the quantiles 
(Fig. \ref{fig3}b for a two-dimensional example) 
shows that in these cases, even though bacteria closest to the source have a lowered uptake, a majority of the bacteria 
that are already above the average uptake in the non-chemotactic case gain access to even higher uptake. Bacteria furthest 
from the source, and below the average uptake level in the non-chemotactic case, see their mean uptake 
decrease even further. Overall this corresponds to an increase of $G$: in higher dimensions, 
chemotaxis can bring the bacteria further away from the ideal free distribution, 
that is, it increases the inequalities among the population. 
The increase or decrease of inequalities of uptake, as revealed by the positive or negative change of $G$, 
may thus depend in detail on the system characteristics embodied in $l$, $L$ and $\alpha$.

\section{Implications for Fitness}

What would be the biological consequences of chemotactic levelling of resources and of uptake rates? 
Uptake of nutrients governs a wide range of bacteriological processes, among which are cell growth and division. 
In particular, the yield of 
biomass per unit nutrient taken up is an increasing function of the available nutrient concentration, a feature which 
has been suggested as selective for the response characteristics of chemotaxis \cite{Celani}. Here we provide a 
brief discussion of how the chemotactically-driven redistribution of resources throughout a population 
can impact on the average growth rate of the population which we consider a measure of fitness.  Continuing the point of view taken in the introduction,
we show that while at the single cell level in a defined resource field chemotaxis may increase fitness, 
this is not necessarily true at the population level.

We compute an average growth rate \(\bar{\mu}\) over the population from the steady-state distributions of 
(\ref{chemotaxis_nodim}), in a model in which the local growth rate \(\mu(x)\) is proportional to the local uptake rate through 
a yield function \(y(c)\): \(\mu(x)=y(c) \;c\) and
\begin{equation}
\bar \mu = \frac{1}{\lambda} \int_0^{\lambda}\! dx\, y(c) c b~.
\end{equation}

\begin{figure}[t]
\includegraphics{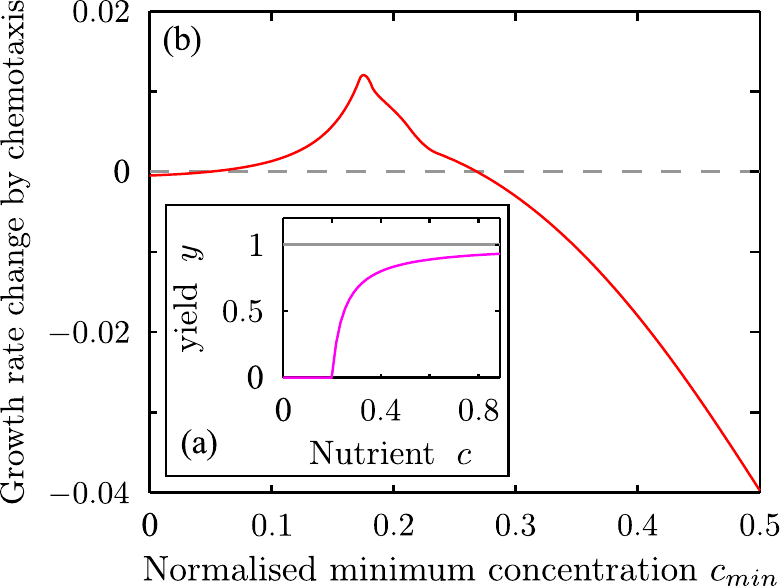}
\caption{(color online). Consequences of uptake levelling on population growth rate.  Using $\alpha=4$, $\lambda=3$ and $s=0.125$ in $d=1$. (a) Yield as a function of nutrient concentration, for $c_{min}=0.2$, $K=0.05$ and $y_0=1$. (b) Change of growth rate as a consequence of chemotaxis for $K=0.05 $ and varying $c_{min}$: if chemotaxis can increase growth for low yield threshold, the uptake leveling makes it less advantageous for higher threshold by diluting the resource.}
\label{fig4}
\end{figure}

Whereas the mean uptake is fixed by the boundary conditions in our problem, this average growth coefficient 
will depend on the distribution of the resource among the bacteria and thus will be modified by chemotaxis. 
In order to capture the increase of \(y\) with \(c\), we consider that \(y=0\) below of threshold concentration 
\(c_{min}\), and adopt a Michaelis-Menten form above it \cite{BookwithDroop}:
\begin{equation}
y(c)=y_0 \frac{(c-c_{min})}{(c-c_{min})+K} \;\; \mbox{for} \;\;c>c_{min}
\end{equation}
with \(K\) a saturation constant of the yield (Fig. \ref{fig4}a).

The threshold concentration \(c_{min}\) can be considered as the limit below which all the uptake 
is directed toward maintenance costs \cite{Tempest}. The relative change of the average growth rate due to chemotaxis (\(\bar{\mu}_{chemo}-\bar{\mu}_{non chemo}\)) is shown in Fig. \ref{fig4}b. 
We observe that for lower values of the threshold concentration \(c_{min}\) the redistributive 
effect increases the number of cells that reach the growth threshold: the population fitness 
becomes higher with chemotaxis than without. However, for higher values of the threshold 
concentration \(c_{min}\), the redistribution of uptake throughout the population leaves more 
cells below the growth threshold: we get the counter-intuitive result that chemotaxis effectively 
lowers population fitness with respect to the non-chemotactic case. This result, which stems from the competition of the bacteria for the 
same resource, shows that in the context of continuous sources, an homogeneously chemotactic population 
could be selected against due to the dilution of a scarce resource resulting from the chemotactic behavior.

\section{Conclusions}

We have shown that the organization of bacteria around localized nutrient sources is fundamentally 
different from that of higher animals due to diffusion of resources and feeders. 
Yet, there are still common characteristics. First, what might be termed `foraging' behavior decreases 
the maximum uptake rate through competition for the resource; in the bacterial case this corresponds to a decrease 
of the maximum of the concentration field with chemotaxis. Second, foraging generates, quite evidently, 
localization and accumulation of the population closer to the resources.  But whereas the conjunction of these two phenomena 
brings Harper's ducks to the ideal free distribution, it may fail in the microbial world: it brings the system 
closer or further from this ideal distribution depending on the spatial dimensionality and parameters capturing 
the strength of chemotaxis, the size of the resources and the distance between them. 
The redistribution of uptake is not without consequences: when the resource is scarce in comparison to the 
metabolic needs, chemotaxis effectively dilutes it and reduce the average population fitness.

The issues addressed here suggest experimental studies of model systems in physical ecology for which 
{\it in situ} measurements of local metabolic activity and nutrient concentration fields are possible.    
Optically-based quantitative measures of photosynthetic activity \cite{photosynthesis}, probes of local oxygen 
concentration \cite{Douarche}, and local mass spectrometry \cite{massspec} are examples of relevant techniques.
Microbial communities in biofilms, sediments \cite{sediments}, and 
algae sustaining a motile population of bacteria around them by releasing oxygen \cite{AdlerTempleton} 
represent interesting systems in which to study 
the distribution of uptake rates.

\section{Acknowledgments}

We are grateful to Wolfram Schultz for bringing Ref. 1 to our attention, and thank O. Croze, T.J. Pedley, 
W. Poon, A. Smith, G. Peng and R. Watteaux for discussions. This work was supported by a Raymond and Beverley Sackler Scholarship (FJP), MinesParisTech (FJP), and the 
European Research Council Advanced Investigator Grant 247333 (REG).

\end{document}